\documentclass[a4paper]{jpconf}
\usepackage{graphicx}
\usepackage{bm}          
\usepackage{amssymb}   
\usepackage{amsmath} 
\usepackage{cleveref}
\begin{document}
\bibliographystyle{iopart-num}

\title{Open-quantum-systems approach to in-medium heavy quarkonium dynamics}

\author{Alexander Rothkopf}

\address{Faculty of Science and Technology, University of Stavanger, 4021 Stavanger, Norway}

\ead{alexander.rothkopf@uis.no}

\begin{abstract}
Heavy quarkonium continues to play a central role in the study of nuclear matter under extremes of temperature and density in relativistic heavy-ion collisions. In this talk I report on recent developments in the theoretical description of quarkonium in-medium dynamics using the open-quantum systems approach. Not only does it provide a clear interpretation of the imaginary part of the complex heavy-quark potential but also reveals that a subtle interplay between screening and wavefunction decoherence is responsible for the melting of heavy quarkonium states.
\end{abstract}

\section{Introduction}

According to the standard model of relativistic heavy-ion collisions (HIC) (for a review see e.g.~\cite{Elfner:2022iae}) the time evolution of matter in the collision center can be roughly divided into three stages. The earliest stages just after the projectile nuclei hit, are characterized by strong coherent color electric and color magnetic fields, the so called glasma. The energy stored in these field is quickly converted into quark and gluon particles which efficiently exchange energy and momentum to form a locally thermal quark-gluon plasma (QGP) after around $t\sim1$fm. This QGP expands and cools over a a range of around $\Delta t\sim5-10$fm, well described by relativistic hydrodynamics. Reaching the transition temperature of strongly interacting matter, predicted by quantum chromodynamics (QCD) at around $155$MeV, colored partons have to come together to form color neutral hadrons at hadronization after which the produced particles freeze out chemically and kinetically before flying out into the surrounding detectors.

Heavy quarkonium, the bound states of a heavy quark and antiquark (charm or bottom) are ideal probes of the quark-gluon plasma created in HICs (for a review see e.g.~\cite{Rothkopf:2019ipj}). Their life cycle spans that of the HIC, with production of $Q\bar Q$ pairs in the earliest instances that may or may not form bound states before finding themselves immersed in the hot QGP. As the QGP cools, quarkonium states are affected by the environment and depending on how deeply they are bound, may survive until hadronization or will dissociate. At hadronization either surviving bound states emerge or formerly dissociated pairs of heavy quarks may find a chance to recombine before flying off into the detectors. Our goal here is to understand from first principles the evolution of a fully formed quarkonium state immersed in the QGP.

Quarkonium dilepton decay channels provide clean signals in experiment \cite{Andronic:2015wma} and an inherent separation of scales simplifies their theory treatment \cite{Brambilla:2004jw}. The fact that the heavy quark mass $m_Q$ is larger than both the environment temperature $T$ and the intrinsic scale of quantum fluctuations in QCD $\Lambda_{\rm QCD}$ allows one to describe the quarks and anti-quarks separately in the effective field theory of non-relativistic QCD (NRQCD). If the quarkonium state is of small extent relative to environment scales it can be described by color singlet and octet wavefunction via potential NRQCD (pNRQCD). While the singlet channel can sustain bound states, the octet channel is associated with a repulsive interaction. Note that the separation in terms of energy scales also translates into a separation of time-scales.

\begin{figure}
\centering
\includegraphics[scale=0.7]{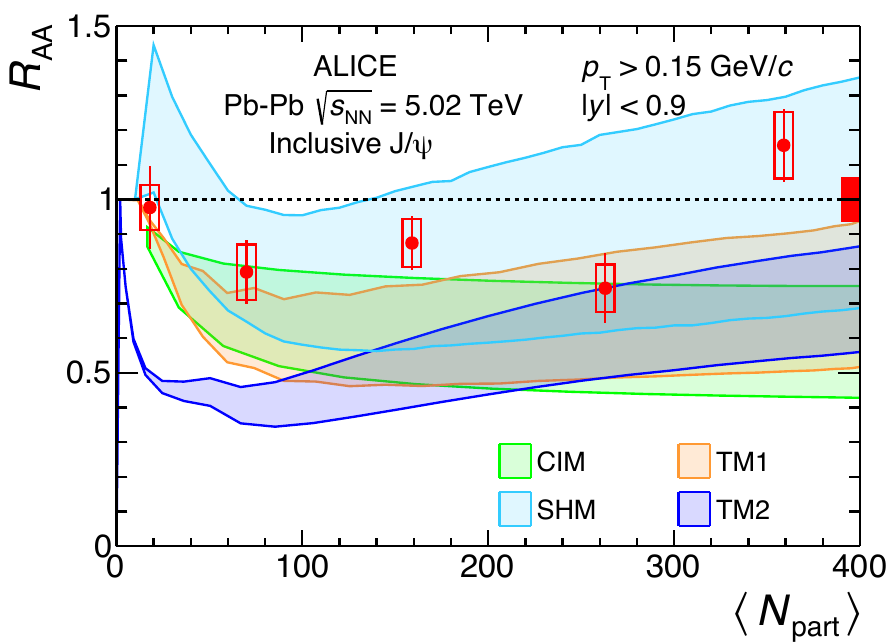}
\includegraphics[scale=0.7]{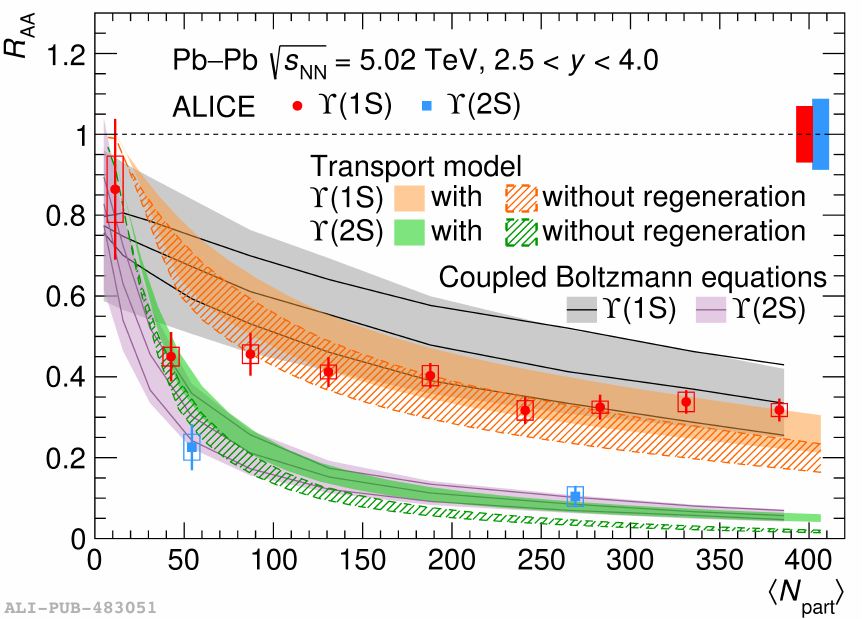}
\caption{Nuclear modification factor $R_{AA}$ for the vector channel (left) charmonium ground state $J/\Psi$ \cite{ALICE:2019nrq} and (right) for the two lowest lying states of bottomonium $\Upsilon(1S)$ and $\Upsilon(2S)$ \cite{ALICE:2020wwx}. Note that multiple phenomenological models, based on significantly different physics processes, are able to capture the data reasonable well, indicating the need for genuine 1st principle insight.}\label{fig:Raa}
\end{figure}

One may ask why a genuine ab-initio understanding of quarkonium dynamics is called for? Phenomenological modelling based on the rate equation (mainly for charmonium) and either the Boltzmann equation or the deterministic Schr\"odinger equation (mainly for bottomonium) have successfully postdicted the nuclear modification factor of several quarkonium species as shown in \cref{fig:Raa}. In the absence of ab-initio insight on the parameters of the models, one however mixes the determination of heavy-quarkonium properties and those of the medium. I.e. inference of QGP properties alone requires first-principles insight into $Q\bar{Q}$ dynamics.

Progress in this direction has been made in the past years through the open quantum systems paradigm \cite{breuer2002theory}, originally developed in condensed matter theory. It represents a general formalism to treat a small probe system immersed in an environment not necessarily in equilibrium. The overall system, consisting of probe and environment is closed and its density matrix $\rho_{\rm tot}$ evolves unitarily $\partial_t\rho_{\rm tot}=-i[H_{\rm tot},\rho]$, according to a hermitian Hamiltonian $H_{\rm tot}^\dagger=H_{\rm tot}$. In order to avoid having to treat the surrounding medium explicitly, we focus on the evolution of the probe d.o.f. and may formally trace out the environment. In turn we obtain the reduced probe density matrix ${\rm Tr}_{\rm medium}[\rho_{\rm tot}]=\rho_{Q\bar Q}$, whose dynamics are in general dissipative, as it remains in contact with the surrounding bath. 

The so called master equation, i.e. the equation of motion that governs $\rho_{Q\bar Q}$ takes on different forms, depending on the separation of three relevant timescales. The environmental relaxation scale $\tau_E$ tells us how quickly perturbations among the environment d.o.f. die down. The intrinsic probe scale $\tau_S\sim 1/|\omega-\omega^\prime|$ is related to the inverse level spacing of the probe and encodes how quickly information is updated in the small system. The third scale is the $Q\bar Q $ relaxation scale $\tau_{\rm rel}$ which encodes how quickly information travels between the medium and the probe system.

For a system in which memory effects can be neglected, i.e. for $\tau_E\ll\tau_{\rm rel}$, so-called Markovian time evolution, it has been proven that the most general master equation can be expressed in the \textit{Lindblad} form
\begin{align}
\frac{d}{dt} \rho_{Q\bar Q}=-i[\bar H_{Q\bar Q},\rho_{Q\bar Q}]+\sum_k\gamma_k\Big( L_k\rho_{Q\bar Q}L_k^\dagger- \frac{1}{2} \{L^\dagger_kL_k,\rho_{Q\bar Q}\}\Big).\label{eq:Lindblad}
\end{align}
Here $\bar H_{Q\bar Q}$ denotes the in-medium probe Hamiltonian with a real-valued potential inducing \textit{coherent dynamics}, while the \textit{Lindblad operators} $L_k$ together with the relaxation rate $\gamma_k$ encode the residual interactions between probe and environment. They can be associated with fluctuations (energy flows into the probe system) and dissipation (energy flows out of the probe system).

\section{Open Quantum Systems approaches for heavy quarkonium}

\begin{figure}
\centering
\includegraphics[scale=0.5]{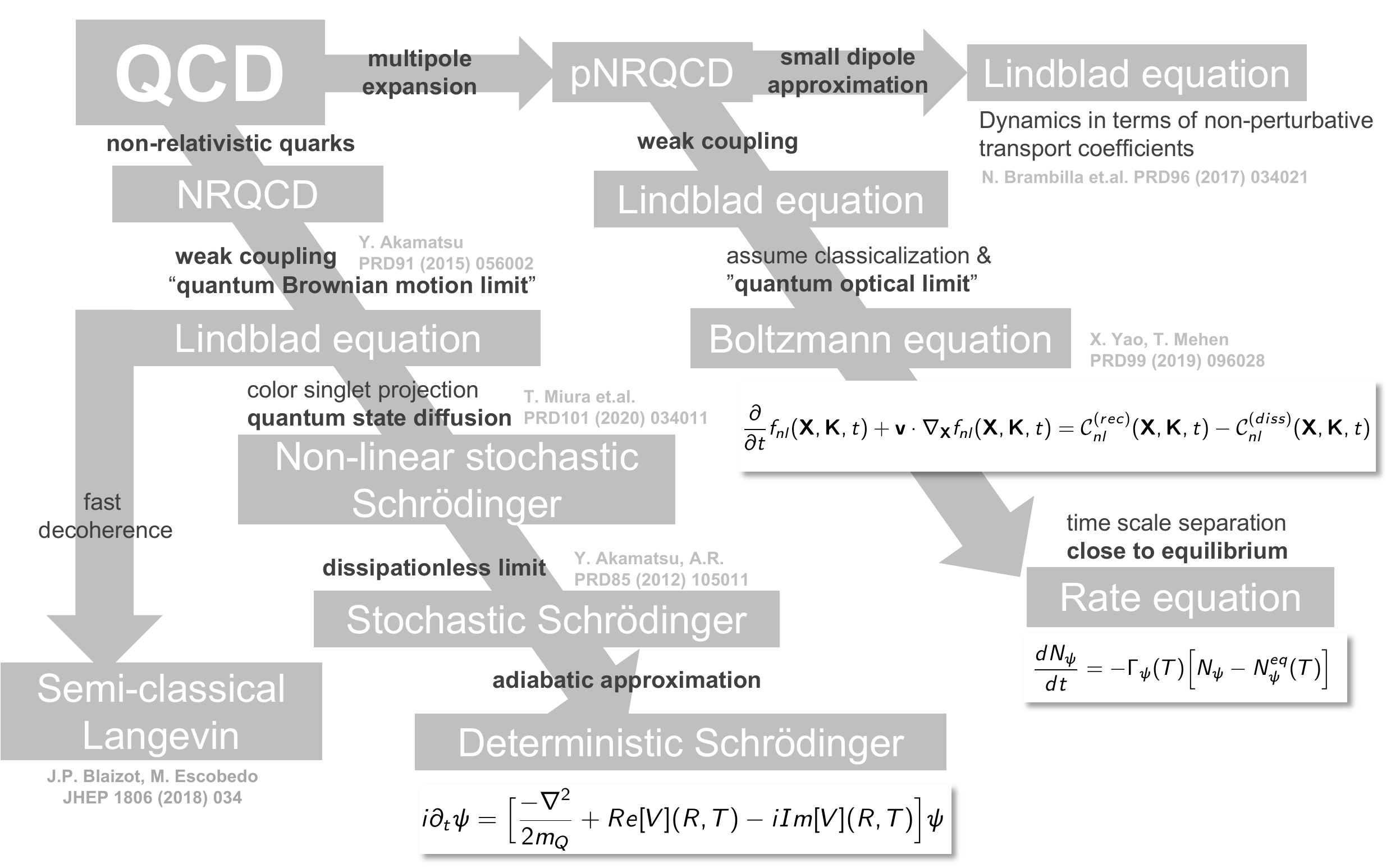}
\caption{The systematic chain of approximations that the open quantum systems approach has built between ab-intio theory QCD, the effective field theories NRQCD and pNRQCD and the phenomenological approaches used in the past. At intermediate steps different genuine quantum descriptions of quarkonium, applicable in regimes with appropriate time-scale separation, have been established.}\label{fig:OQSforQQbar}
\end{figure}

Over the past seven years the open quantum systems approach has been used to derive from first principles QCD several Lindblad equations for heavy quarkonium (for a review see e.g. \cite{Akamatsu:2020ypb}), applicable in different regimes of scale separation (see \cref{fig:OQSforQQbar}). One part of the community uses the effective field theory pNRQCD as starting point, which, when evaluated in the weak-coupling limit, gives rise to a Lindblad equation \cite{Yao:2018nmy}. In the \textit{quantum optical limit} and under the assumption of fast classicalization this equation be reduced to the Boltzmann equation \cite{Yao:2020xzw}. Assuming that the system is close to equilibrium, it may further be reduced to the rate equation used in phenomenological modelling. This establishes a systematic chain of approximations with well defined ranges of validity from QCD to phenomenology. To overcome the weak-coupling approximation, another research branch focuses on very tightly bound quarkonium states (Coulombically bound) \cite{Brambilla:2016wgg}, in which the dynamics of the system can be conveniently captured in two non-perturbative transport coefficients $\kappa$ and $\gamma$ \cite{Brambilla:2020qwo}. The former is nothing but the heavy-quark diffusion constant and the latter amounts to a correction term for the coherent dynamics. Both quantities are being actively investigated using lattice QCD simulations (see e.g. \cite{Altenkort:2020fgs}).

Another active research direction within the community considers the open-quantum systems approach for quarkonium based on the effective field theory NRQCD, in which the quark and antiquark remain as individual entities. If one considers a weakly-coupled hot environment, one can argue that the information within the quarkonium probe refreshes much more slowly than among the medium d.o.f., leading ref.~\cite{Akamatsu:2014qsa} to establish a \textit{quantum Brownian motion} Lindblad equation. It is known in the open quantum systems community that the dynamics of a Lindblad equation can always be \textit{unravelled}, i.e. reexpressed in terms of an ensemble of wavefunctions, evolving under a non-linear stochastic Schr\"odinger equation \cite{Miura:2019ssi}. This for the first time allowed to connect models with such a non-linear equation of motion (see e.g. \cite{Katz:2015qja}) to first principles QCD. Interestingly one can now carry out further controlled approximations to arrive at other phenomenological models previously deployed in the literature. In the dissipationless limit, the non-linear stochastic Schr\"odinger equation reduces to a linear one \cite{Akamatsu:2011se,Kajimoto:2017rel}. And if in turn one uses the adiabatic approximation, a deterministic Schr\"odinger equation with an complex valued potential ensues (see e.g. \cite{Krouppa:2017jlg}). It is interesting to note that in this NRQCD based approach it is the real- and imaginary part of the complex interquark potential, originally discovered in ref.~\cite{Laine:2006ns} and currently investigated on the lattice (see e.g. \cite{Rothkopf:2011db,Burnier:2014ssa,Burnier:2015tda,Bala:2021fkm}), that characterizes the evolution.

If one focuses on systems where the quarkonium decoheres very quickly \cite{Blaizot:2018oev}, one may also go over to a semi-classical description \cite{Blaizot:2017ypk} in terms of a Langevin equation for the quarks and antiquarks, which offers insight into e.g. the mechanism of recombination.

\section{The Quantum Brownian Motion Lindblad equation}

Let me focus on the quantum Brownian motion Lindblad equation for heavy quarkonium, as it has revealed that two distinct processes compete in the evolution of the bound state in a hot medium (see discussion in \cite{Kajimoto:2017rel}). In the standard form of \cref{eq:Lindblad} we get that
\begin{align}
\bar H_{Q\bar Q}=\frac{{\bf p}^2}{m_Q}+V_{Q\bar Q}({\bf r}), \quad V_{Q\bar Q}({\bf r})=-\alpha_S\frac{e^{-m_D|{\bf r}|}}{|{\bf r}|}+c={\rm Re}[V_{\rm EFT}({\bf r})],
\end{align}
which indicates that Debye-screening of the interactions with Debye mass $m_D$ will affect the binding of the quark antiquark pair. This mechanism of destabilization was known since the seminal works of ref.~\cite{Matsui:1986dk}. Let us take a closer look at the Lindblad operators themselves. Since behind the scenes, each interaction of the quarkonium system with a medium gluon leads to a momentum exchange and a color rotation we expect these phenomena to be represented in the $L$'s, which indeed carry a momentum and color label
\begin{align}
L_{{\bf k},a}=&\sqrt{\frac{D({\bf k})}{2}}\Big[1-\frac{{\bf k}}{4m_QT}\cdot \Big(\frac{1}{2}{\bf P}_{\rm CM}+{\bf p}_{\rm rel}\Big)\Big]e^{i{\bf k}\cdot{\bf r}/2}\big(T^a\otimes 1\big)\\
&-\sqrt{\frac{D({\bf k})}{2}}\Big[1-\frac{{\bf k}}{4m_QT}\cdot \Big(\frac{1}{2}{\bf P}_{\rm CM}+{\bf p}_{\rm rel}\Big)\Big]e^{i{\bf k}\cdot{\bf r}/2}\big(1\otimes T^a\big).
\end{align}
In NRQCD quark and antiquark are separate entities and hence we find two terms, with opposite signs. The color rotations on either one are implemented via the Gell-Mann matrices $T^a$. The momentum independent term $1$ in the square brackets encodes the physics of fluctuations, i.e. it allows energy from the medium to enter the quarkonium system and heat up. The momentum dependent terms on the other hand are associated with dissipation, which allows quarkonium to shed energy back to the medium. Obviously only the correct interplay between the two will allow the system to thermalize. The overall strength of the interaction with the medium is governed by a function $D({\bf k})$, which, as it turns out, is intimately connected in coordinate space with the imaginary part of the complex interquark potential $D({\bf r})\sim{\rm Im}[V_{\rm EFT}({\bf r})]$. 

The function $D({\bf r})$ exhibits a peaked behavior in coordinate space, which allows us to identify a new scale in the system, the medium correlation length $\ell_{\rm corr}$. With this new scale comes a new mechanism for the destabilization of heavy quarkonium: \textit{wavefunction decoherence}. It is thus the interplay of screening and decoherence that governs the survival of heavy quarkonium.

As temperature increases, the screening length $r_s=1/m_D$ reduces. As long as it is larger than the interquark distance, the gluons exchanged between the constituent quark antiquark pair are not impacted. As the screening length shrinks and reaches the characteristic extent of the bound state, gluon exchange is impeded and eventually binding cannot be sustained.

The medium correlation length $\ell_{\rm corr}$ encodes over which distances color rotations act coherently. As long as $\ell_{\rm corr}$ is larger than the size of quarkonium, color rotation act in sync on the quark and antiquark constituent, turning a color singlet state into another color singlet state. As the correlation length diminishes, color rotations may act independently on the quark or antiquark, leading to an increasing probability to realize a color octet configuration that is not attractively bound. This can happen at a temperature where screening has not yet impacted the exchange of gluons between the constituents of the quarkonium particle.

\begin{figure}
\centering
\includegraphics[scale=0.5]{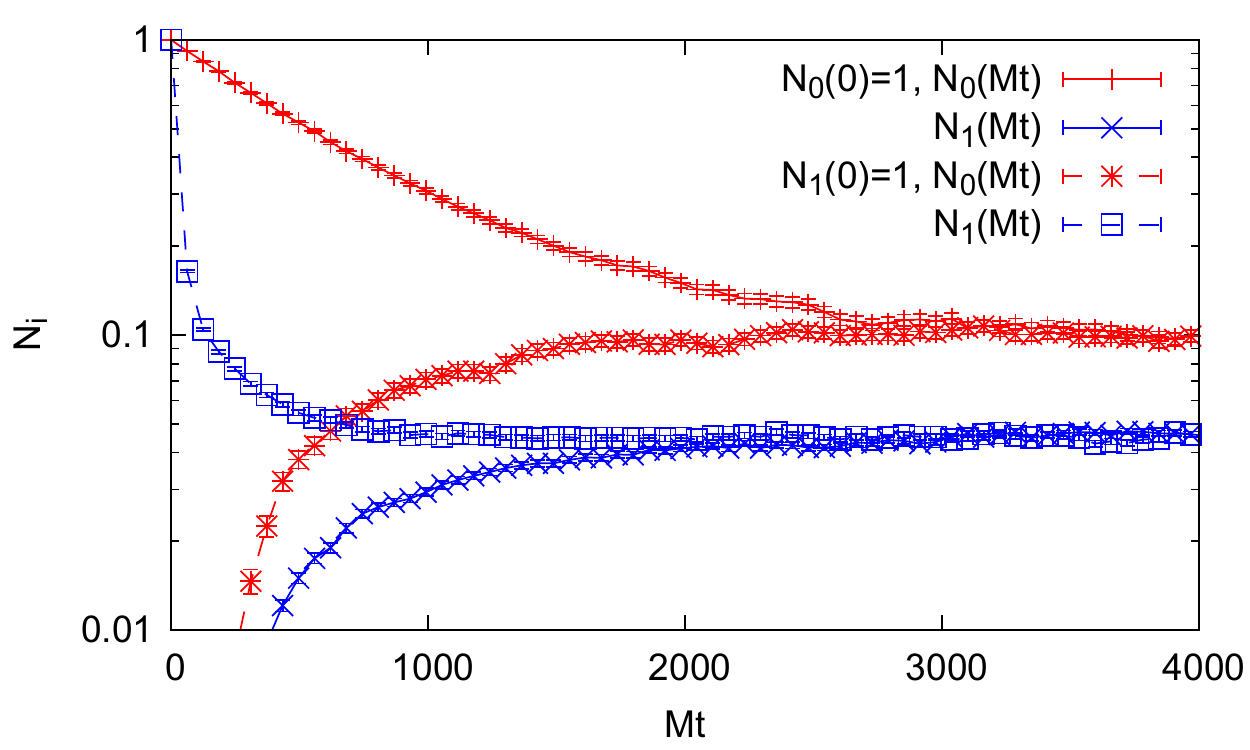}
\includegraphics[scale=0.5]{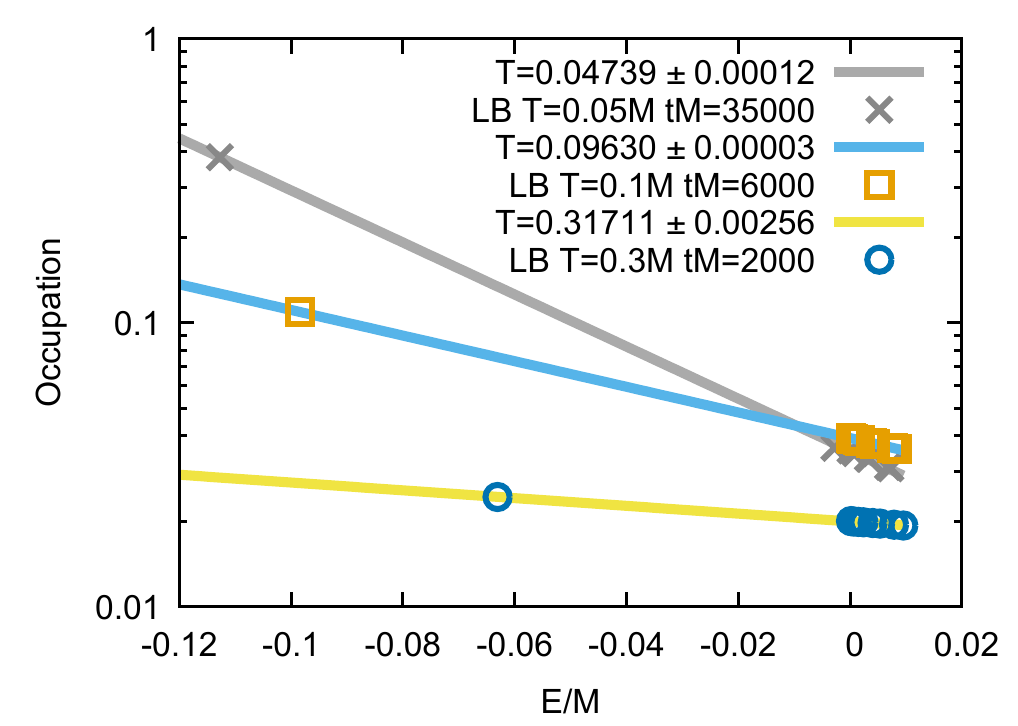}
\caption{First results of the quantum Brownian motion Lindblad equation in one dimension. (left) The evolution of a quarkonium vacuum bound state analog in 1d evolving in a thermal medium \cite{Miura:2019ssi}. Red data denote ground state survival probabilities, blue data the first excited state. Independent of initial conditions the two simulations shown approach a universal steady state. (right) The distribution of states in the late-time steady state, plotted against energy at different temperatures \cite{Alund:2020ctu}. }\label{fig:NumRes}
\end{figure}

The quantum Brownian motion Lindblad equation was first implemented in a one-dimensional setting in ref.~\cite{Miura:2019ssi} using quantum state diffusion and in ref.~\cite{Alund:2020ctu} via a direct simulation of the master equation. In both cases the projection onto the singlet sector was taken apriori. In the left panel of \cref{fig:NumRes} we show a central result of these studies, where the one-dimensional analog of a quarkonium vacuum bound state at initial time was evolved in a medium with finite screening and correlation length. The red data represent the survival probability of the ground state, while the blue points describe the first excited state. Two simulations are shown, one where we start with $100\%$ ground and one with $100\%$ excited state at initial time. At late time both simulations approach the same steady state, which is independent of the initial conditions. To ascertain whether the steady state is related to a genuine themalized system we show in the right panel of \cref{fig:NumRes} three spectra from simulations at late times, where we plot the survival probability against the energy of each state. The data points clearly align on a straight line when plotted on logarithmic scale, which agrees with the exponential falloff expected from the thermal Boltzmann weight. A fit to the inverse slope reveals values for the inverse slope, which are very close to the temperature of the medium in which we made the quarkonium particle evolve. 

These results are very encouraging, as they confirm that the quantum Brownian motion Lindblad equation is able to thermalize a quarkonium state in a genuine quantum fashion. So far this is the only quantum approach on the market that has been demonstrated to correctly thermalize quarkonium. In order for thermalization to occur fluctuations and dissipation need to be in balance. Our simulations show that dissipative effects, contrary to intuition, can have a significant effect on the ground state survival probability at early times. This insight was obtained by a direct comparison of the full Lindblad dynamics and the reduced dynamics in the dissipationless limit. Simulations including the color singlet and octet d.o.f. explicitly have been carried out by the Osaka group recently \cite{Akamatsu:2021vsh} and the extension of the simulations to three dimensions is work in progress.

\section{Conclusion}

The open quantum systems approach has provided a new perspective on quarkonium dynamics in relativistic heavy-ion collisions. It has made possible to derive systematic chains of approximations that for the first time link phenomenological models with first principles QCD using the effective field theories NRQCD or pNRQCD at intermediate steps. The quantum Brownian motion Lindblad equation derived from NRQCD revealed the interplay of screening and wavefunction decoherence as two relevant mechanism for the dynamic destabilization of quarkonium in a hot environment and was shown to correctly thermalize the quark antiquark pair with its surroundings in a genuine quantum fashion.

\section{Acknowledgments}
A.R. is supported by the Research Council of Norway under the FRIPRO Young Research Talent grant 286883.

\section*{References}

\bibliography{iopart-num}

\end{document}